\documentclass[11pt]{article}
\usepackage{amsfonts}
\usepackage{amssymb}
\usepackage{amsmath}
\usepackage{latexsym}
\begin{document}

\title{\bf Thermal rectification in quantum graded mass systems}

\author{Emmanuel Pereira\\
{\small Departamento de F\'{\i}sica-ICEx, UFMG, CP 702, 30.161-970 Belo Horizonte MG, Brazil}\\
{\small e-mail:  emmanuel@fisica.ufmg}}

\date{\today}

\maketitle

\begin{abstract}
We show the existence of thermal rectification in the graded mass
quantum chain of harmonic oscillators with self-consistent
reservoirs. Our analytical study allows us to identify the
ingredients leading to the effect. The presence of rectification in
this effective, simple model (representing graded mass materials,
systems that may be constructed in practice) indicates that
rectification in graded mass quantum systems may be an ubiquitous
phenomenon. Moreover, as the classical version of this model does
not present rectification, our results show that, here,
rectification is a direct result of the quantum statistics.

\vspace{.5cm} \noindent {\bf PACS} number(s): 05.70.Ln; 05.40.-a;
44.10.+i

\vspace{.5cm}


\end{abstract}

\vspace{1cm}

A fundamental challenge in statistical physics is the derivation of
macroscopic phenomenological laws of thermodynamic transport from
the underlying microscopic Hamiltonian systems. However, after
decades, a first-principle derivation of the Fourier's law of heat
conduction, for instance, is still a puzzle \cite{BLRb}. Many works
have been devoted to the theme \cite{LLP}, most of them by means of
computer simulations. But, in these problems where the central
question involves the convergence or divergence of thermal
conductivity, sometimes there is a significant difficulty to arrive
at precise conclusions from numerical results \cite{GS}. Thus, the
necessity of analytical studies together with the huge complexity of
the associated nonlinear dynamical systems led to several works
considering approximative schemes or simplified models \cite{Dhar} -
since the pioneering work of Debye, the microscopic models to
describe heat conduction are mainly given by systems of anharmonic
oscillators, leading to problems without analytical solutions. Many
works involve, e.g., the use of approximations such as Boltzmann
equations, master equations for effective models, the analysis of
Green-Kubo formula, etc. An example of simple (effective) model that
can be analytically studied is the classical or quantum harmonic
chain of oscillators with self-consistent reservoirs. It has been
proposed a while ago \cite{BRV,VR}, but it is always revisited
\cite{BLL,DR}. In such model, each site of the chain is coupled to a
reservoir; the first and last sites are coupled to ``real" thermal
baths, while the inner reservoirs only mimic the absent anharmonic
interactions. The self-consistent condition means that there is no
heat flow between an inner reservoir and its site in the steady
state: the inner baths act only as a phonon scattering mechanism,
such as the on-site anharmonic potential in more elaborated systems.
This model, the classical and also the quantum version, in
opposition to a standard harmonic system, obeys the Fourier's law
\cite{BLL}.

In this scenario of intensive study of the heat mechanism, problems
involving the possibility to control the heat flow by constructing
thermal nano-instruments such as thermal diodes, and even
transistors, thermal gates and memories, have recently attracted
considerable theoretical \cite{Casati,BHu,BLi2} and experimental
\cite{Chang} interest. A thermal diode, or rectifier, is a device in
which the magnitude of the heat current changes as we invert the
device between two thermal baths. That is, in a thermal diode, heat
flows preferably in one direction. There are some analytical
attempts to explain the phenomenon of thermal rectification and/or
design a diode by using simple models (see e.g. the spin-boson
nanojunction \cite{SN}, \cite{WS}, and the billiard system
\cite{EM}) but, again, most of the works consider numerical
computations \cite{Casati,BHu,BLi2}. It is worth to recall the
extensive work of B. Li and collaborators: e.g. thermal
rectification in asymmetric graphene ribbons is studied by using
molecular dynamics simulations in ref. \cite{BLi95}; also in carbon
nanocone structures \cite{BLi93}; at silicon-amorphous polyethylene
interfaces \cite{BLi92}; in carbon nanotube intramolecular junctions
\cite{BLiPRB07}, etc. A commonly used design of diodes is given by
the sequential coupling of two or three chains with different
anharmonic potentials \cite{Casati,BHu,BLi2}. It is frequently
studied, but it is also criticized due to the difficulty to be
constructed  in practice \cite{BHu}. Recently, Chang et al.
\cite{Chang}, considering a different procedure, built a nanoscale
thermal rectifier in an experimental work: they use a graded
material, namely, nanotubes externally and inhomogeneously
mass-loaded with heavy molecules, a system with asymmetric-axial
thermal conductance. Graded materials are also considered in other
numerical studies of diodes \cite{BLi,C}, and the interesting
results made some authors suggest that thermal rectification is
guaranteed by the use of graded materials with anharmonic
interactions.

Having in mind the investigation of this suggested relation between
graded (anharmonic) materials and rectification, in a recent paper
\cite{PLPRE} we study the classic harmonic chain of oscillators with
self-consistent reservoirs (CSC). We show the absence of thermal
rectification for the CSC model with graded masses or graded
interparticle interactions by using the method developed in
refs.\cite{PF1}, \cite{PF2}. In \cite{S2008}, Segal obtains similar
results by using the approach of \cite{DR}. We, and other authors
\cite{BRV,BLL}, understand the SC models as effective anharmonic
models (the inner reservoirs mimic the absent anharmonic
potentials), but it is not a consensus. Anyway, even if we discard
any relation with anharmonicity, these results show, at least, that
the onset of Fourier's law does not guarantee thermal rectification
in asymmetric classic models - we recall that the Fourier's law does
not hold in standard harmonic systems (i.e., harmonic models with
reservoirs at the boundaries only), but it holds in the SC models.

Still searching for the mechanism behind the rectification in graded
mass systems, we recall that, at low temperatures, quantum effects
may introduce significant changes. The thermal conductivity for the
quantum self-consistent harmonic chain (QSC), for example, depends
on temperature \cite{DR,FLP}, in opposition to the CSC thermal
conductivity \cite{BLL}. Thus, searching for possible quantum
effects in the mechanism of thermal rectification, in a quite recent
work \cite{PLJPA}, we turn to the inhomogeneous QCS and, in the
linear response regime, i.e., for the chain submitted to a very
small gradient of temperature and considering only linear
corrections in computations with changes in the temperature, we show
the absence of thermal rectification despite quantum effects in the
conductivity. However, we may still ask about the possibility of a
rectification for the quantum system submitted to a large gradient
of temperature or, at least, for some effect beyond linear
corrections. It is, indeed, a very important question: in
ref.\cite{CMP}, the authors show that there is a significative
rectification in some chaotic billiard systems ``provided the
temperatures (of the two sides of the system) are strongly
different..."

In the present work, with the focus on the onset of rectification in
inhomogeneous systems, we revisit the graded mass QSC model, and
study it by using an analytical approach valid for any temperature
gradient, i.e., beyond the linear response regime. For this quantum
model, we show that, in opposition to the behavior of its classical
version, there is a thermal rectification. We offer an explanation
for this difference: in the quantum version, the expression for the
heat flow (\ref{Fl}) involves a distribution for the phonon
frequencies (the frequencies in the system) that depends on
temperature (\ref{phonon}), leading to a mix of temperature,
frequencies and also particle masses (see eqs.(\ref{Fl}, \ref{GW})).
It does not happen in the classical system - see the comments for
the high temperature limit at the end of this manuscript. This
intricate mix of temperature, masses and frequencies makes the heat
current change as we invert the graded mass system between two
thermal reservoirs. The presence of rectification in this quite
simple, say, bare model is of considerable interest: it indicates
the generality of such phenomenon in the experimentally realizable
class of graded mass systems, i.e., it indicates that rectification
is not related to intricate interactions in the microscopic systems.
Moreover, as far as we know, this work is the first demonstration of
a ``purely" quantum thermal rectifier, in the sense that,  for such
model, the rectification is a direct result of the quantum
statistics: i.e., rectification is absent in the classical version
of the model.

Let us introduce the QSC model. We use a Ford-Kac-Mazur approach, as
detailed presented in ref.\cite{DR}, in order to describe the
quantum system and its time evolution to the steady state. Here, all
the baths connected to the chain are modeled as mechanical harmonic
systems, with initial coordinates and momenta determined by some
statistical distribution. Then, we solve the quantum dynamics given
by Heisenberg equations, take the stochastic distribution for the
initial coordinates of the baths, as well as the limit $t \to
\infty$, and obtain the expression for the heat flow in the steady
state.

The Hamiltonian of our system, a chain (W) with harmonic
interparticle and on-site potentials, with each site connected to a
bath $(B)$, also with harmonic interactions, is given by
\begin{eqnarray} \label{Hquant}
\mathcal{H} &=& \mathcal{H}_W + \sum_{i=1}^N \mathcal{H}_{B_i} +
\sum_{i=1}^N X_W^T V_{B_i} X_{B_i},
\\
\mathcal{H}_S &=& \frac{1}{2} \dot{X}_S^T M_S \dot{X}_S +
\frac{1}{2} X_S^T \Phi_S X_S , \nonumber
%\\
%\mathcal{H}_{B_i} &=& \frac{1}{2} \dot{X}_{B_i}^T M_{B_i}
%\dot{X}_{B_i} + \frac{1}{2} X_{B_i}^T \Phi_{B_i} X_{B_i} , \nonumber
\end{eqnarray}
where $S=W$ or $B_{i}$; $M_W$, $M_{B_i}$ are the particle--mass
diagonal matrices for the chain and baths; $\Phi_W$ and $\Phi_{B_i}$
are symmetric matrices describing the interparticle and on-site
harmonic interactions; and $V_{B_i}$ gives the interaction between
the $i$-site and its bath (more details below). We have, for each
part, $X=\left[ X_1,X_2,\ldots, X_{N_s}\right]^T$, where $X_r$ is
the position operator of the $r$-th particle; $\dot{X}=M^{-1}P$,
where $P_l$ is the momentum operator of the $l$-th particle. Of
course, it follows that $[X_r,P_l]=i \hbar \delta_{r,l}$. Finally,
the dynamics is given by the Heisenberg equations
\begin{eqnarray} \label{dynquant}
M_{W} \ddot{X}_{W} &=& - \Phi_{W} X_{W} - \sum_i V_{B_i}^T X_{B_i} ,
\\
M_{B_i} \ddot{X}_{B_i} &=& - \Phi_{B_i} X_{B_i} - \sum_i V_{B_i}^T X_{W} .
\nonumber
\end{eqnarray}

The formulas for the heat currents inside the chain and from each
reservoir to the chain are related to $\langle X_{W} \dot{X}_{W}^T
\rangle$ and $\langle X_{B} \dot{X}_{W}^T \rangle$. Before
presenting the formulas, let us give a very short resume of their
derivation. To find the expressions, we turn to the Heisenberg
equations (\ref{dynquant}), treat the equations of the baths as
linear inhomogeneous equations, and plug these solutions into the
equations for the chain. Then, we take the average over the initial
conditions of the baths, which are assumed to be distributed
according to equilibrium phonon distributions with properly chosen
temperatures, determined such that the self-consistent condition
holds, that is, we must take the temperatures of the inner baths
such that there is no heat flow between an inner bath and its site.
We reach the steady state by taking the limit $t \to \infty$. For
technical reasons we still take $t_0 \to -\infty$, and consider the
Fourier transform of $t$. We note that, by plugging the solutions of
the baths back into the equations of motion for the system, we get a
quantum Langevin equation. Remember that, for our model, all
particles are connected to heat reservoirs that are taken to be
Ohmic. The coupling strength to the reservoirs is controlled by the
dissipation constant $\zeta$ defined by an expression involving the
matrix $V$ presented in the Hamiltonian above (see ref.\cite{DR}).

The expression for the heat flow from the $l$-th reservoir to the
chain is given by
\begin{eqnarray} \label{Fl}
\mathcal{F}_l &=& \sum_{m=1}^N \zeta^2 \int_{-\infty}^{+\infty} d
\omega \ \omega^2 \left| \left[ G_W(\omega) \right]_{l,m}
\right|^2 \times \\
& & \times \ \frac{\hbar \omega}{\pi} \left[ f(\omega,T_l) -
f(\omega,T_m) \right] ,
\nonumber
\end{eqnarray}
where $\zeta$ is the dissipation constant;
\begin{equation} \label{GW}
[G_W(\omega)]^{-1} =  - \omega^2 M_W + \Phi_W - \sum_l
\Sigma_l^{+}(\omega) ,
\end{equation}
the matrix $\Sigma_l^{+}$ above has only one non-vanishing element:
$(\Sigma_l^{+})_{l,l} = i \zeta \omega$; $f(\omega,T_l)$ is the
phonon distribution for the $l$-th bath
\begin{equation} \label{phonon}
f(\omega,T_l) = 1/[ \exp(\hbar \omega/K T_l) - 1 ];
\end{equation}
the variable $\omega$ in the expressions above comes from the
Fourier transform, namely, $ \widetilde{X}(\omega) = (2\pi)^{-1}
\int_{-\infty}^{+\infty} d t \ X(t) \ e^{i \omega t}$. We stress
that a detailed derivation of the heat flow formula (\ref{Fl}) is
presented in ref.\cite{DR}.

The heat flow in the chain may be given by $\mathcal{F}_{1}$, i.e.,
by the heat flow from the first reservoir, together with the
self-consistent condition
$\mathcal{F}_{2}=\mathcal{F}_{3}=\ldots=\mathcal{F}_{N-1}=0$, which
means that the inner reservoirs do not inject energy into the system
(the energy comes from the first reservoir, passes trough the chain
and goes out to the last reservoir). In such condition, we have
$\mathcal{F}_{1}=-\mathcal{F}_{N}$. To investigate the presence or
absence of thermal rectification, we need to examine the heat
current in a chain with reservoirs at temperatures $T_{1}, T_{2},
\ldots, T_{N}$, where $T_{1}$ and $T_{N}$ are given, and $T_{2},
\ldots, T_{N-1}$ are determined by the self-consistent condition
($\mathcal{F}_{2}=0, \ldots, \mathcal{F}_{N-1}=0$); and also in the
same chain with inverted thermal reservoirs, precisely, with the
baths taken at temperatures $T'_{1}, T'_{2}, \ldots, T'_{N}$, where
$T'_{1}=T_{N}$, $T'_{N}=T_{1}$, and $T'_{2}, \ldots, T'_{N-1}$ are
determined by the self-consistent condition ($\mathcal{F}'_{2}=0,
\ldots, \mathcal{F}'_{N-1}=0$). In the case of absence of thermal
rectification, we shall obtain $\mathcal{F}_{1}+\mathcal{F}'_{1}=0$.
That is what we will examine.

The determination of $G_{W}(w)$, the matrix given by eq. (\ref{GW})
and necessary to study $\mathcal{F}_{l}$, may be a very difficult
task -- we know a precise solution for some specific cases: e.g.,
for a homogeneous next-neighbors interparticle interaction and all
particles with the same mass $M_W=mI$ \cite{DR}; and also for the
case of particles with alternate masses, i.e. $m_j=m_1$ for $j$ odd,
and $m_j=m_2$ for $j$ even \cite{FLP}. Hence, we first restrict our
interparticle interaction $\Phi$ to the homogeneous nearest-neighbor
case, i.e., $\Phi = -\Delta$, the lattice Laplacian, and study the
equations for the smallest possible chain: $N=3$ (it is, in fact, a
cell of a large chain). Then, we show that a graded mass
distribution leads to a thermal rectification. After that, we argue
to show that similar relations extend to $N=4$, and so on.

Thus, for $N=3$, we have
\begin{equation}
G_{W}^{-1}(w) = \left[
  \begin{array}{ccc}
    R_{1,1} & -1 & 0 \\
    -1 & R_{2,2} & -1 \\
    0  & -1 & R_{3,3}\\
  \end{array}
\right],
\end{equation}
where $R_{k,k}\equiv g^{-1}_{k,k} = 2-w^{2}M_{k}-i\zeta w$. To have
graded masses, we take $M_{1}<M_{2}<M_{3}$. And using the notation
$c \equiv \zeta^{2}\hbar/\pi$, $A_{lm} \equiv
|(G_{W}(w))_{l,m}|^{2}$, $f_{k} \equiv f(w,T_{k})$, we get (from
eq.(\ref{Fl}))
\begin{eqnarray*}
\mathcal{F}_{1} &=& c\int dw ~ w^{3}A_{12}[f_{1}-f_{2}] + ~ c\int
dw ~ w^{3}A_{13}[f_{1}-f_{3}],\\
 \mathcal{F}_{3} &=& c\int dw ~
w^{3}A_{31}[f_{3}-f_{1}] + ~ c\int dw ~ w^{3}A_{32}[f_{3}-f_{2}].
\end{eqnarray*}
It is easy to see that $A_{jk}=A_{kj}$. By taking the inverse of
$G_{W}^{-1}$ above, we have
\begin{eqnarray}
A_{12} &=& [(2-w^{2}M_{3})^{2}+\zeta^{2}w^{2}]/|det
G_{W}^{-1}|^{2},\\
A_{32} &=& [(2-w^{2}M_{1})^{2}+\zeta^{2}w^{2}]/|det
G_{W}^{-1}|^{2}.\nonumber
\end{eqnarray}
Note that, by using $\mathcal{F}_{1}=-\mathcal{F}_{3}$, from the
equations above we obtain (as expected) $\mathcal{F}_{2}=0$. For the
chain with inverted reservoirs, we obtain the expressions for
$\mathcal{F}'_{1}$ and $\mathcal{F}'_{3}$ similar to those for
$\mathcal{F}_{1}$ and $\mathcal{F}_{3}$ above, but with the change
$T_{1}\leftrightarrow T_{3}$, i.e., with the replacement
$f_{1}\leftrightarrow f_{3}$, and $f_{2}\rightarrow f'_{2}$. The
question is: do we have $\mathcal{F}_{1} + \mathcal{F}'_{1}=0$?

From $\mathcal{F}_{1}=-\mathcal{F}_{3}$ (i.e.,
$\mathcal{F}_{2}=0$) and $\mathcal{F}'_{1}=-\mathcal{F}'_{3}$
(i.e., $\mathcal{F}1_{2}=0$, we obtain
\begin{eqnarray}
\int dw ~ w^{3}A_{12}[f_{1}-f_{2}]  &=& \int
dw ~ w^{3}A_{32}[f_{2}-f_{3}],\label{I}\\
 \int dw ~
w^{3}A_{12}[f_{3}-f'_{2}] &=& \int dw ~ w^{3}A_{32}[f'_{2}-f_{1}].
\label{II}
\end{eqnarray}
To get the absence of rectification, we must have
$\mathcal{F}_{1}+\mathcal{F}'_{1}=0$, i.e.,
\begin{equation}
\int dw ~ w^{3}A_{12}[f_{1}-f_{2}]  = \int dw ~
w^{3}A_{12}[f'_{2}-f_{3}],\label{III}
\end{equation}
or, which comes directly from eqs.(\ref{I},\ref{II},\ref{III}) (and
also from $\mathcal{F}_{3}+\mathcal{F}'_{3}=0$)
\begin{equation}
\int dw ~ w^{3}A_{32}[f_{3}-f_{2}]  = \int dw ~
w^{3}A_{32}[f'_{2}-f_{1}].\label{IV}
\end{equation}
That is, if we have eqs.(\ref{I},\ref{II},\ref{III}), then
eq.(\ref{IV}) follows; similarly, given
eqs.(\ref{I},\ref{II},\ref{IV}), we have eq.(\ref{III}).

Let us use the short notation
$$
F_{1}(T_{k}) = \int dw ~ w^{3}A_{12}f_{k}, ~~ F_{2}(T_{k}) = \int
dw ~ w^{3}A_{32}f_{k},
$$
and similarly for $F_{j}(T'_{2})$.
 From eqs.(\ref{I}) and (\ref{II}) we
obtain
\begin{eqnarray}
F_{1}(T_{2}) + F_{2}(T_{2}) &=& F_{1}(T_{1}) + F_{2}(T_{3}),\label{I*}\\
F_{1}(T'_{2}) + F_{2}(T'_{2}) &=& F_{2}(T_{1}) +
F_{1}(T_{3}).\label{II*}
\end{eqnarray}
%\begin{eqnarray}
%\int dw ~w^{3}(A_{32}+A_{12})f_{2}  = \int
%dw ~w^{3}(A_{12}f_{1}+A_{32}f_{3}),\label{I*}&&\\
% \int dw ~w^{3}(A_{32}+A_{12})f'_{2}  = \int dw
%~w^{3}(A_{12}f_{3}+A_{32}f_{1}),\label{II*}&&
%\end{eqnarray}
We write the condition for absence of rectification (\ref{III}) as
\begin{equation}
F_{1}(T_{2}) + F_{1}(T'_{2}) = F_{1}(T_{1}) +
F_{1}(T_{3}),\label{III*}
\end{equation}
%\begin{equation}
%\int dw ~ w^{3}(A_{12}f'_{2} + A_{12}f_{2})  = \int dw ~
%w^{3}(A_{12}f_{1} + A_{12}f_{3}),\label{III*}
%\end{equation}
or, which comes from these equations,
\begin{equation}
F_{2}(T_{2}) + F_{2}(T'_{2}) = F_{2}(T_{1}) +
F_{2}(T_{3}).\label{IV*}
\end{equation}
%\begin{equation}
%\int dw ~ w^{3}(A_{32}f'_{2} + A_{32}f_{2})  = \int dw ~
%w^{3}(A_{32}f_{1} + A_{32}f_{3}).\label{IV*}
%\end{equation}

Let us briefly give an idea of our analysis before turning to the
mathematical expressions. It follows that any function $F_{j}(T)$
above, and also the sum $F_{1}(T) + F_{2}(T)$, is a monotone,
strictly increasing, convex function of $T$ (it is easily verified
by studying its derivative $d/dT$). Hence, given $T_{1}$ and
$T_{3}$, the value of $T_{2}$ is univocally determined by
eq.(\ref{I*}), and the value of $T'_{2}$ by eq.(\ref{II*}).
Moreover, $F_{1}(T)$ is bigger and increases faster than $F_{2}(T)$,
for $M_{1}<M_{2}<M_{3}$. In short, we have three different equations
for $T_{2}$ and $T'_{2}$, that are determined by the two first
equations; as the third one may not be derived (it is independent)
from the first and second equations, we shall have rectification,
i.e. the third equation will not be always satisfied. For linear
equations in $T$ (that is the case for the approximations of high
temperature or linear response regime, as we show ahead), the system
of two variables and three equations has a solution only if it is
linearly dependent - more details ahead.

Now we make explicit, i.e., we give the mathematical details of the
analysis described above for the eqs.(\ref{I*}, \ref{II*},
\ref{III*}). For simplicity, we will consider small temperatures
(and so, also small changes in $T$), but, of course, we will go
beyond linear corrections. Since $F$ ($F_{1}$ and $F_{2}$) is an
analytic function of $T$ (for $T>0$), we write, for some $T_{0}\neq
0$, $|T_{0}-T|$ small,
\begin{eqnarray}
F_{1}(T) &=& F_{1}(T_{0}) + F'_{1}(T_{0})(T-T_{0}) +
\frac{F''_{1}(T_{0})}{2}(T-T_{0})^{2} \nonumber \\
&=& \left[F_{1}(T_{0}) - F'_{1}(T_{0})T_{0} +
\frac{F''_{1}(T_{0})}{2}T_{0}^{2}\right] + \left[F'_{1}(T_{0}) -
2T_{0}\frac{F''_{1}(T_{0})}{2}\right]T +
\frac{F''_{1}(T_{0})}{2}T^{2} \nonumber \\
&\equiv & a_{0} + a_{1}T + a_{2}T^{2} \label{F1},
\end{eqnarray}
that is, we discard corrections up to $\mathcal{O}((T-T_{0})^{3})$.
We have (at least for small $T_{0}$) $a_{1},a_{2}>0$. The same
follows for $F_{2}(T)$,
\begin{equation}
F_{2}(T) = b_{0} + b_{1}T + b_{2}T^{2} \label{F2}.
\end{equation}
Hence, introducing eqs.(\ref{F1}, \ref{F2}) in eq.(\ref{I*}), we
get, for small $T_{1}$ and $T_{3}$,
\begin{equation}
T_{2} = \frac{-(a_{1}+b_{1}) \pm \sqrt{(a_{1}+b_{1})^{2} +
4(A_{1}+B_{3})(a_{2}+b_{2})}}{2(a_{2}+b_{2})} \label{T2},
\end{equation}
where
\begin{equation}
A_{1}\equiv a_{1}T_{1} + a_{2}T_{1}^{2} , ~~~~ B_{3}\equiv
b_{1}T_{3} + b_{2}T_{3}^{2}.
\end{equation}
Of course, only $T_{2}>0$ makes sense, and so, $T_{2}$ is determined
by the positive expression in eq.(\ref{T2}) above. Similarly, for
$T'_{2}$ we get, from eq.(\ref{II*}),
\begin{equation}
T'_{2} = \frac{-(a_{1}+b_{1}) \pm \sqrt{(a_{1}+b_{1})^{2} +
4(A_{3}+B_{1})(a_{2}+b_{2})}}{2(a_{2}+b_{2})} \label{T'2},
\end{equation}
where
\begin{equation}
A_{3}\equiv a_{1}T_{3} + a_{2}T_{3}^{2} , ~~~~ B_{1}\equiv
b_{1}T_{1} + b_{2}T_{1}^{2} .
\end{equation}
Again, we consider the positive expression for $T'_{2}$.

Now we check if eq.(\ref{III*}) is verified or not, that means
absence or existence of rectification respectively. Namely, we check
if it is true the equality
\begin{equation}
(a_{1}T_{2} + a_{2}T_{2}^{2}) + (a_{1}{T'}_{2} + a_{2}{T'}_{2}^{2})
= (a_{1}T_{1} + a_{2}T_{1}^{2}) + (a_{1}T_{3} + a_{2}T_{3}^{2})
\label{ret}.
\end{equation}
To make the investigation, we introduce the (positive) expressions
for $T_{2}$ (\ref{T2}) and $T'_{2}$ (\ref{T'2}) in equation above.
By taking $T_{1}=T$, $T_{3}=T+\epsilon$, with small $\epsilon$, and
using
$$
\sqrt{1 + x} = 1 + \frac{1}{2}x - \frac{1}{8}x^{2} +
\mathcal{O}(x^{3}),
$$
we easily verify that eq.(\ref{ret}) is not satisfied. E.g., by
taking in eq.(\ref{ret}) the terms involving only $\epsilon$ and
$\epsilon^{2}$ (not those with $T$, $T\epsilon$, etc), we get, for
the L.H.S. and R.H.S. of eq.(\ref{ret}),
\begin{eqnarray*}
{\rm L.H.S.} &=& a_{1}\epsilon +
\frac{a_{1}}{a_{1}+b_{1}}(a_{2}+b_{2})\epsilon^{2} +
(a_{2}b_{1}-a_{1}b_{2})\frac{(b_{1}^{2}+a_{1}^{2})}{(a_{1}+b_{1})^{3}}\epsilon^{2}
, \\
{\rm R.H.S.} &=& a_{1}\epsilon + a_{2}\epsilon^{2} .
\end{eqnarray*}
Note that there is an equality up to first order in $\epsilon$, and
so, as we have already recalled \cite{PLJPA}, we do not see
rectification in the linear regime. But, considering the corrections
beyond the linear approximation, the difference is evident, which
shows the existence of rectification.

For transparency, let us also give more details of our analysis  for
some situations of absence of rectification. It happens, e.g., in
the following cases. (I) For the system with homogeneous mass
distribution (obviously): if $M_{1}=M_{2}=M_{3}$, then
$A_{12}=A_{32}$ (and so, eq.(\ref{III*}) is obtained by the sum of
eqs.(\ref{I*}) and (\ref{II*})). (II) In the high temperature
regime: in such case, as shown in ref.\cite{DR}, the quantum system
behaves as the classical one, where there is no rectification (as
already shown in refs.\cite{PLPRE} and \cite{S2008}). To see this
result from our previous equations, note that for large $T$ we have
$w/[\exp(\hbar w/KT) - 1] \rightarrow KT/\hbar$, and so,
eqs.(\ref{I*}), (\ref{II*}), and (\ref{III*}) become linear
equations in $T_{2}$, $T'_{2}$, $T_{1}$ and $T_{3}$. It is easy to
see that they are linearly dependent (and so, eq.(\ref{III*}) comes
from eqs.(\ref{I*}) and (\ref{II*})). (III) In the linear response
regime (already shown in ref.\cite{PLJPA} and commented above):
again, as described in ref.\cite{DR}, for small temperature
gradients we have
\begin{equation*}
\mathcal{F}_{k} = c\int dw ~ w^{3} \frac{\partial f}{\partial
T}(w,T) \sum_{m=1}^{3}A_{km}(T_{k}-T_{m}),
\end{equation*}
where $T$ above is some average temperature. The expression above
leads us to the same case of (II) (i.e., we have linear equations
in $T_{2}$, $T'_{2}$, etc).

Now we turn to larger chains. For $N=4$, we have the equations
$\mathcal{F}_{1} = -\mathcal{F}_{4}$ and $\mathcal{F}_{2}=0$,
$\mathcal{F}_{3}=0$ (the self-consistent equation). But, as
$\mathcal{F}_{1} + \mathcal{F}_{2} + \mathcal{F}_{3} +
\mathcal{F}_{4} = 0$ (it may be easily verified from the expressions
for $\mathcal{F}$), if we take $\mathcal{F}_{1}= - \mathcal{F}_{4}$
and $\mathcal{F}_{2}=0$, then $\mathcal{F}_{3}=0$ follows (i.e., it
is not a new equation; and vice-versa: $\mathcal{F}_{2}=0$ and
$\mathcal{F}_{3}=0$ leads to $\mathcal{F}_{1}= - \mathcal{F}_{4}$).
Given $T_{1}$ and $T_{4}$, these two equations (e.g.,
$\mathcal{F}_{1}= - \mathcal{F}_{4}$ and $\mathcal{F}_{2}=0$)
determines $T_{2}$ and $T_{3}$. The same follows for the system with
inverted reservoirs ($T'_{2}$ and $T'_{3}$). Hence, the equation for
the absence of rectification $\mathcal{F}_{1} + \mathcal{F}'_{1} =
0$ is an extra equation involving variables already determined, and
the analysis follows as before (for $N=4$ and any other $N$). We
emphasize that the study of the dependence of the rectification on
the system size (not presented here) is of importance: recall that,
in the two-segment diode, the rectification vanishes as the system
size goes to infinite.

We note that, for a system with three sites, if we take a mass
distribution with $M_{1}\neq M_{3}$ (e.g., $M_{1}=M_{2}\neq M_{3}$,
or $M_{1}\neq M_{2}=M_{3}$) we still have rectification. For larger
$N$, the analysis for the graded system is ``automatic", but it may
be intricate for any other mass distribution. It is an interesting
(and difficult) problem to investigate which inhomogeneousity in the
mass distribution is enough to guarantee the rectification.

We still recall that, for a standard harmonic system, i.e., for the
harmonic chain with reservoirs at the boundaries only (in other
words, turning off the inner reservoirs, say, the ``schematic
anharmonicity''), we have (see ref.\cite{DR}),
$$
\mathcal{F}_{1} = c\int dw ~ w^{3}A_{1N}(f_{1}-f_{N}),
$$
and similarly for $\mathcal{F}'$, with the replacement
$f_{1}\leftrightarrow f_{N}$; i.e.,
$\mathcal{F}_{1}=-\mathcal{F}'_{1}$, and so, there is no
rectification.

To conclude, our results show the existence of thermal rectification
in the graded mass QSC, a quite simple model. It indicates that the
presence of some phonon scattering mechanism, or some effective
anharmonicity, is enough to guarantee the thermal rectification in
an inhomogeneous quantum model. We stress the importance of the
quantum nature of the reservoirs that provides an energy mixing
(\ref{phonon}). It is probable that some other intricate mechanism,
such as real anharmonic potentials, may provide such mix in
classical models, as indicated by numerical studies \cite{BLi}.
Finally, we recall that graded mass models are not only theoretical
systems: they may be constructed in practice \cite{Chang}. In short,
diodes of graded mass systems sound to be ubiquitous (and
experimentally reliable) structures.

\section*{Acknowledgments}
We are in debt to the referee of a previous (unpublished) work to
point out a mistake in our tentative to show the absence of thermal
rectification for the QSC model.  We also thank the reviewers of the
present manuscript for several comments that improved its
presentation. Work supported by CNPq and Fapemig (Brazil).


\baselineskip 0.5cm
\begin{thebibliography}{}

\bibitem{BLRb} F. Bonetto, J. L. Lebowitz,
L. Rey-Bellet, Mathematical Physics 2000, Imperial College Press
(2000) pp. 128-150.

\bibitem{LLP} S. Lepri, R. Livi, A. Politi, Phys. Rep. 377 (2003) 1-80.

\bibitem{GS} O. V. Gendelman, A. V. Savin, Phys. Rev. Lett. 92 (2004) 074301.

\bibitem{Dhar} A. Dhar, Adv. Phys. 57 (2008) 457.

\bibitem{BRV} M. Bosterli, M. Rich, W.M. Visscher, Phys. Rev. A 1
(1970) 1086.

\bibitem{VR} W.M. Visscher, M. Rich, Phys. Rev. A 12 (1975) 675.

\bibitem{BLL} F. Bonetto, J.L. Lebowitz, J. Lukkarinen, J. Stat.Phys 116 (2004) 783.

\bibitem{DR} A. Dhar, D. Roy, J. Stat. Phys. 125 (2006) 801.

\bibitem{Casati} M. Terraneo, M. Peyrard, G. Casati, Phys. Rev. Lett. 88, (2002) 094302.

\bibitem{BHu} B. Hu, L. Yang, Y. Zhang, Phys. Rev. Lett. 97 (2006) 124302.

\bibitem{BLi2} B. Li, L. Wang, G. Casati, Phys. Rev. Lett. 93 (2004) 184301.

\bibitem{Chang} C.W. Chang, D. Okawa, A. Majumdar, A. Zettl, Science 314 (2006) 1121.

\bibitem{SN} D. Segal, A. Nitzan, Phys. Rev. Lett.94 (2005) 034301.

\bibitem{WS} L.-A. Wu, D. Segal, Phys. Rev. Lett.102 (2009) 095503.

\bibitem{EM} J.P. Eckmann, C. Mej\'{\i}a-Monasterio, Phys. Rev. Lett.
97 (2006) 094301.

\bibitem{BLi95} N. Yang, G. Zhang, B. Li, Appl. Phys. Lett. 95 (2009) 033107.

\bibitem{BLi93} N. Yang, G. Zhang, B. Li, Appl. Phys. Lett. 93 (2008) 243111.

\bibitem{BLi92} M. Hu, P. Keblinski, B. Li, Appl. Phys. Lett. 92 (2008) 211908.

\bibitem{BLiPRB07} G. Wu, B. Li, Phys. Rev. B 76 (2007) 085424.



\bibitem{BLi} N. Yang, N. Li, L. Wang, B. Li, Phys. Rev. B 76 (2007) 020301 (R).

\bibitem{C} G. Casati, Nat. Nanotechnol. 2 (2007) 23.


\bibitem{PLPRE} E. Pereira,  H. C. F. Lemos,  Phys. Rev. E 78  (2008) 031108.

\bibitem{PF1} E. Pereira, R. Falcao, Phys. Rev. E 70 (2004) 046105.

\bibitem{PF2} E. Pereira, R. Falcao, Phys. Rev. Lett. 96 (2006) 100601.


\bibitem{S2008} D. Segal,   Phys. Rev. E 79,  (2009) 012103.

\bibitem{FLP} A. Francisco Neto, H.C.F. Lemos, E. Pereira, Phys.
Rev. E 76 (2007) 031116.

\bibitem{PLJPA} E. Pereira,   H. C. F. Lemos,
  J. Phys. A: Math. Theor. 42 (2009) 225006.

\bibitem{CMP} G. Casati, C. Mej\'{\i}a-Monasterio, T. Prosen, Phys. Rev. Lett.
98 (2007) 104302.



\end{thebibliography}
\end{document}